\documentstyle[multicol,aps,epsf]{revtex} 
\tighten
\newcommand{\PostScript}[7]{
\begin{figure}[H]
\begin{center}
\leavevmode
\epsfysize=#1cm
\vspace{#2cm}
\epsfbox{#3}
\par
\parbox{#5cm}{
\vspace{#4cm}
\caption[figure]{\renewcommand{\baselinestretch}{1} \small \normalsize #6}
\label{#7}}
\end{center}
\end{figure}
}

\begin{document}

\title{Quantum dots in magnetic fields: Phase diagram and 
broken symmetry of the Chamon-Wen edge}

\author{S.M. Reimann, M. Koskinen, M. Manninen and B.R. Mottelson$^*$}      
\address{\it Department of Physics, University of Jyv\"askyl\"a,
FIN-40351 Jyv\"askyl\"a, Finland\\
$^*$NORDITA, Blegdamsvej 17,DK-2100 Copenhagen, Denmark}

\medskip

\maketitle

\begin{abstract}
Quantum dots in magnetic fields are studied 
within the current spin density functional formalism
avoiding any spatial symmetry restrictions of the solutions.
We find that the maximum density droplet reconstructs 
into states with broken internal symmetry: 
The Chamon-Wen edge co-exists with a modulation of the charge density 
along the edge. The phase boundaries between the 
polarization transition, the maximum density droplet and its reconstruction 
are in agreement with recent experimental results. 
\end{abstract}

\begin{multicols}{2}
\narrowtext

Quantum dots are small electron islands made 
by laterally confining the two-dimensional electron gas in a 
semiconductor heterostructure. Such nano-sized  systems  
attracted much interest since the techniques in their 
fabrication developed far beyond 
mesoscopic dimensions~\cite{kouwenhoven:review}. Vertical quantum dots 
can nowadays be made so small that they 
show atom-like behavior~\cite{tarucha}: 
shell structure and Hund's rules determine the electronic properties.

Much experimental effort focussed 
on systematically mapping the magnetic field dependence of 
the chemical potential obtained from 
single-electron capacitance spectroscopy~\cite{ashoori}. 
As a bias is applied to the gates, single electrons  
tunnel into the quantum dot when its chemical potential $\mu (N,B)$
(which depends on the number of 
confined electrons $N$ and the magnetic field strength $B$)
equals the Fermi energy in one gate electrode.
First experiments along these lines were performed by 
Ashoori~{\it et al.}~\cite{ashoori} and later Klein {\it et al.}~\cite{klein}. 
Recently, Oosterkamp {\it et al.}~\cite{oosterkamp}
systematically extended the measurements to 
stronger fields $B$ and larger sizes $N$.
Cusps and steps in $\mu (N,B)$ were found to clearly  
separate different ranges of magnetic fields.
From a comparison to results of exact diagonalization studies~\cite{exact} 
these patterns were identified with phase transitions in the droplet: 
they occur at magnetic fields for which the ground-state charge 
distribution of the dot changes, defining sharp phase boundaries. 
The points at which a complete polarization of the electrons occurs 
mark the beginning of the so-called Maximum Density Droplet (MDD) phase.
This new state suggested by McDonald, Yang and Johnson~\cite{mcdonald} 
is a homogeneous droplet in which the density is approximately constant at the 
maximum value 
$\rho _0 =(2\pi l_B^2)^{-1}$ that can be reached in the lowest 
Landau level. ($\ell _B = \sqrt{\hbar /eB}$ is the magnetic length).  
In the spin-polarized MDD the electrons occupy adjacent orbitals with 
consecutive angular momentum. 
This compact occupation of states maximizes the electron density. 
The stability of the MDD is determined by a competition 
between the kinetic and external confinement contributions 
to the total energy, and the Coulomb repulsion of the electrons.
The former would favor the MDD structure up to infinite fields:
with increasing $B$, the droplet would decrease in radius such that  
close to the dot center it could maintain 
a density corresponding to filling factor one in the bulk limit~\cite{foot3}.
This, however, is inhibited by Coulomb repulsion: 
At a sharply defined transition point, the charge density distribution of the 
droplet reconstructs~\cite{mcdonald,cw}. Chamon and Wen~\cite{cw} 
found from Hartree-Fock calculations that at the edge of a quantum Hall liquid
with bulk filling factor one, a stripe or ring of electron density  
breaks off from the homogeneous bulk or the MDD phase.

In this Letter, we show that the so-called Chamon-Wen  
edge actually is a ring 
of nearly localized electrons surrounding the MDD (see Fig.~\ref{fig3} below).
Moreover, we find that the calculated phase boundaries are 
in good qualitative agreement with 
recent experimental results~\cite{oosterkamp}.

As exact diagonalization techniques~\cite{exact} 
are limited to small particle numbers, mean-field methods are needed 
to simplify the complicated many-body problem. We apply the so-called
current spin density functional theory (CSDFT)~\cite{vignale}
including gauge fields in the energy functional.
In contrast to the HF calculations by Chamon and Wen~\cite{cw}
or recent applications of CSDFT~\cite{lubin,lipparini,steffens}
we avoid any spatial symmetry restrictions of the mean field solution.

As a basic model for a quantum dot one usually considers 
$N$ interacting electrons of effective mass $m^*$
confined in a two-dimensional harmonic trap.
A homogeneous magnetic field ${\bf B} = B {\bf e}_z$ is applied
perpendicular to the $x$-$y$-plane in which the electrons 
are confined by the external potential $V=m^*\omega ^2r^2/2$.
In a symmetric gauge, the external vector potential acting on the electrons
is ${\bf A}=B/2 (-y,x)$. For the details of CSDFT we refer to the 
work of Vignale and Rasolt~\cite{vignale} and recall here only 
the most important steps.
The self-consistent Kohn-Sham type equations in CSDFT read
$$
\left[{{\bf p}^2\over 2m^*} + {e\over 2m^*} \left( 
{\bf p} \cdot {\boldmath \cal A}~+~ {\boldmath \cal A}\cdot
{\bf p} \right)
+ {\cal V}_{\delta }\right ]\Psi _{i\delta}=\varepsilon _{i\delta }
\Psi _{i\delta}\nonumber
$$
(We have dropped the arguments $\bf r$ for simplicity).
The index $i$ labels the eigenstates with spin 
$\delta = (\uparrow, \downarrow )$.
We defined effective vector and scalar potentials
${\boldmath \cal A}:= {\bf A}+ {\bf A}_{\rm xc}$
and ${\cal V}_{\delta }:= {(e^2/ 2m^*)}A^2 + V_{\delta}
+ V_H+ V_{{\rm xc}\delta }.$
Here, $V_H$ is the ordinary Hartree potential and $V_{\delta }
= V +(-) g^*\mu _B B/2$
is the external potential, including the Zeeman energy. 
($\mu _B=e\hbar /(2m_e)$ is the Bohr magneton).
The exchange-correlation vector and scalar potentials are 
$$
e{\bf A}_{\rm xc} = {1\over \rho}
\left\{{\partial\over \partial y} {\partial [\rho e_{\rm xc}(\rho _{\delta },\gamma )]
\over \partial {\gamma }}~ ,~ 
- {\partial\over \partial x} {\partial [\rho e_{\rm xc}(\rho _{\delta },{\gamma })]
\over \partial {\gamma } }\right\}
$$
and
$$
V_{{\rm xc}\delta }= {\partial [\rho e_{\rm xc}(\rho _{\delta },{\gamma })]\over 
\partial 
\rho _{\delta }}-{e\over \rho } {\bf j}_p\cdot{\bf A}_{\rm xc}~,  
$$
where $\rho $ is the particle density $\rho = \rho_{\uparrow }+\rho_{\downarrow }$ with
$\rho _{\delta } = \sum _i | \Psi _{i\delta } |^2$~. 
The paramagnetic current density is given by
$
{\bf j}_p =  -{{i\hbar / (2m^*)}}\sum _{i\delta}
[\Psi_{i\delta}^*\nabla \Psi_{i\delta}
-\Psi_{i\delta}\nabla \Psi_{i\delta}^*]
$
and the real current density equals
${\bf j}= {\bf j}_p+ {(e / m^*)}{\bf A} \rho $.
In essence and as applied in
CSDFT, the exchange-correlation energy depends on the so-called vorticity 
${\gamma } = \left. \nabla \times ({{\bf j}_p / \rho })
\right | _z ~$ of the wave function. 
In the bulk the total current density must be zero, and thus
$\gamma =-eB/m^*$. It is this relation that allows to use 
the interpolation formulae for the 
exchange energy per particle $e_{\rm xc}$ for the homogeneous bulk 
in CSDFT by replacing $B\rightarrow (m^*/e)|\gamma |$.
Making use of the local spin density approximation (LSDA),
the exchange energy per particle $e_         {\rm xc}$ is parametrized in terms 
of the local particle density $\rho $, the 
spin polarization $\xi = (\rho _{\uparrow } - \rho _{\downarrow})/\rho $ and
the filling factor $\nu = 2\pi \hbar \rho /m^* |\gamma |$.
We used the expression 
$\label{exc}
e_{\rm xc}(\rho ,\xi,\nu)
=e_{\rm xc}^{\infty }(\rho ) e^{-f(\nu)}
+e_{\rm xc}^{\rm TC}(\rho ,\xi) (1-e^{-f(\nu )})
$, where $f(\nu)=1.5\nu+7\nu^4$. This form interpolates between the infinite
magnetic field limit 
$e_{\rm xc}^{\infty }(\rho ) = -0.782 \sqrt{2\pi \rho } e^2/4\pi \epsilon_0 
\epsilon$
and zero field limit $e_{\rm xc}^{\rm TC}(\rho ,\xi)$, for which we use the 
Tanatar-Ceperly~\cite{tanatar} functional and generalize it to intermediate 
polarizations~\cite{sdwprl}.
For $\nu<0.9$, the interpolation in 
$e_{\rm xc}(\rho ,\xi,\nu)$ follows closely the results of
Fano and Ortolani~\cite{fano} for polarized electrons in the lowest
Landau level, and saturates quickly to the zero field result for $\nu>1$. 

The LSDA has been shown to describe very accurately addition 
spectra~\cite{ashoori,klein,oosterkamp} in weak 
or zero fields~\cite{steffens,isspic}.
Here we concentrate on the polarization transition and beyond.
We use a plane wave basis 
to solve the single-particle Kohn-Sham equations self-consistently.
The practical computational techniques which we found necessary 
to obtain converged solutions of the CSDFT mean field equations 
are described in Ref.~\cite{koskinen}.  
For the material parameters we choose
the typical GaAs values for the effective mass
$m^*=0.067m_e$, the 
dielectric constant $\varepsilon = 12.4$ and the reduced Land\' e g-factor 
$g^*=0.44$. (This yields an effective Bohr radius of $a_B^*=9.79$nm.)
The strength of the external confinement is set to 
$\hbar \omega =4.192~N^{-1/4}$meV~. 
For different $N$, this convention keeps the average electron density 
in the droplet approximately constant, in this case roughly 
corresponding to a (two-dimensional) Wigner-Seitz radius $r_s=2 a_B^*$.

As a first example we study $N=20$ electrons confined  
in the above specified GaAs dot.
At a field of about $2.3$T, only one minority spin is left 
in the dot center. The droplet 
becomes completely polarized for fields larger than about 2.4T.
Shortly after the polarization point the MDD is formed.
Its density profile has perfect azimuthal symmetry 
and is nearly constant inside the droplet corresponding to  
filling factor one in the bulk:
all electrons are in the lowest Landau level and
all single-particle states 
are occupied by one spin down electron with successive 
angular momenta $m=0,1,...,N-1$~\cite{foot4}.
The total orbital angular momentum then is $M={1\over 2}N(N-1)$, which 
yields $M=160$.
As in the inner regions of the MDD the filling factor 
equals one, the density rises with increasing magnetic field, and the 
whole droplet decreases in radius.
For $N=$20 the MDD is stable for fields up to $B\stackrel{<}{\sim}$2.9T, 
where the reconstruction of the charge density 
distribution begins.
The density distribution of the MDD at $B=2.9$T is shown in the 
upper panel of Fig.~\ref{fig1}. 
\PostScript{10}{0}{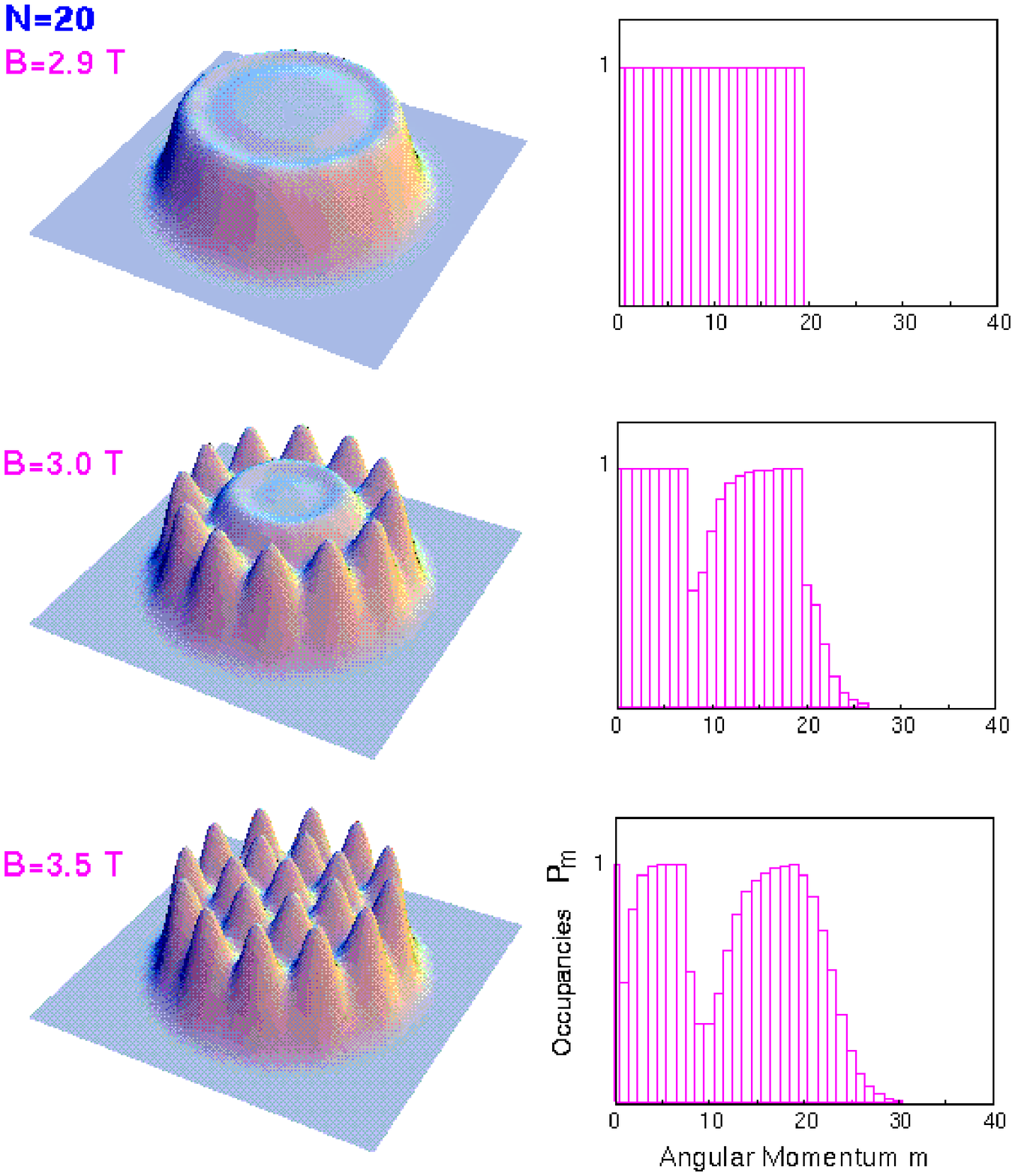}{0.5}{14}{\normalsize {\it Left:}
Self-consistent densities for a 20-electron GaAs quantum dot  
shortly before edge reconstruction at $B=2.9$T ({\it top}), forming 
the broken-symmetry Chamon-Wen edge at $3.0$T ({\it middle}) 
and being fully reconstructed at $3.5$T~({\it bottom}). 
{\it Right:} Angular momentum occupancies $P_m$ as defined 
in the text}{fig1}
We can gain information about the angular momentum occupancy in the droplet 
by projecting the Kohn-Sham single-particle wave functions 
$| \Psi _{i\downarrow }\rangle $ on Fock-Darwin~\cite{fd} 
states $| nm\rangle $ with radial quantum number $n$ and 
good angular momentum $m$. Summing over all occupied states $i\le N$,
we obtain the total angular momentum occupancy 
$P_m=\sum _{i,n} | \langle nm | \Psi _{i\downarrow }
\rangle | ^2$ with $i\le N$. 
(In strong fields, only $n=0$ states give a significant contribution.)
The values of $P_m$ are plotted with bars for each $m$-value 
in the right column of Fig.~\ref{fig1}.
As expected, for the MDD all angular momenta up to $N-1$ have occupancy one.

Increasing the magnetic field effectively compresses the electron states,
such that states with constant $m$ decrease in radius.
Consequently, the density of the MDD would continuously increase with $B$.
This, however, is inhibited by the Coulomb interaction. The droplet  
thus re-distributes~\cite{cw} its density over the dot area, 
taking advantage of moving electrons from lower to higher angular 
momentum states and leaving some partly unoccupied states for $m < N-1$. 
Thus, $M > N(N-1)/2$ and the Chamon-Wen edge is formed~\cite{cw}.
The corresponding density and angular momenta at $B=$3.0T are shown in the 
middle panel of Fig.~\ref{fig1}. After reconstruction, the {\it MDD has thrown 
out a ring of separate lumps of charge density}, with each lump containing 
one electron and having a radius somewhat larger than the magnetic length 
$\ell _B$.
This Chamon-Wen edge with broken rotational symmetry 
is located at a distance of $\sim 2\ell _B$ from the inner part of
the droplet, which continues to form a (smaller) 
MDD with filling factor one in the dot center. 
With the reconstruction of the droplet, 
fractions of angular momenta between $m=8$ and $m=14$ are moved 
to values above $m=19$. The total angular momentum now is raised to
$M=205.13$. (Note that for the {\it internally} broken symmetry of the mean 
field solution, the total angular momentum can take fractional values.)

The total current ${\bf j}(x,y)$ is plotted as a vector diagram 
in Fig.~\ref{fig2}.
It shows vortices which are located around the density maxima 
along the edge.
\noindent
\PostScript{4}{0}{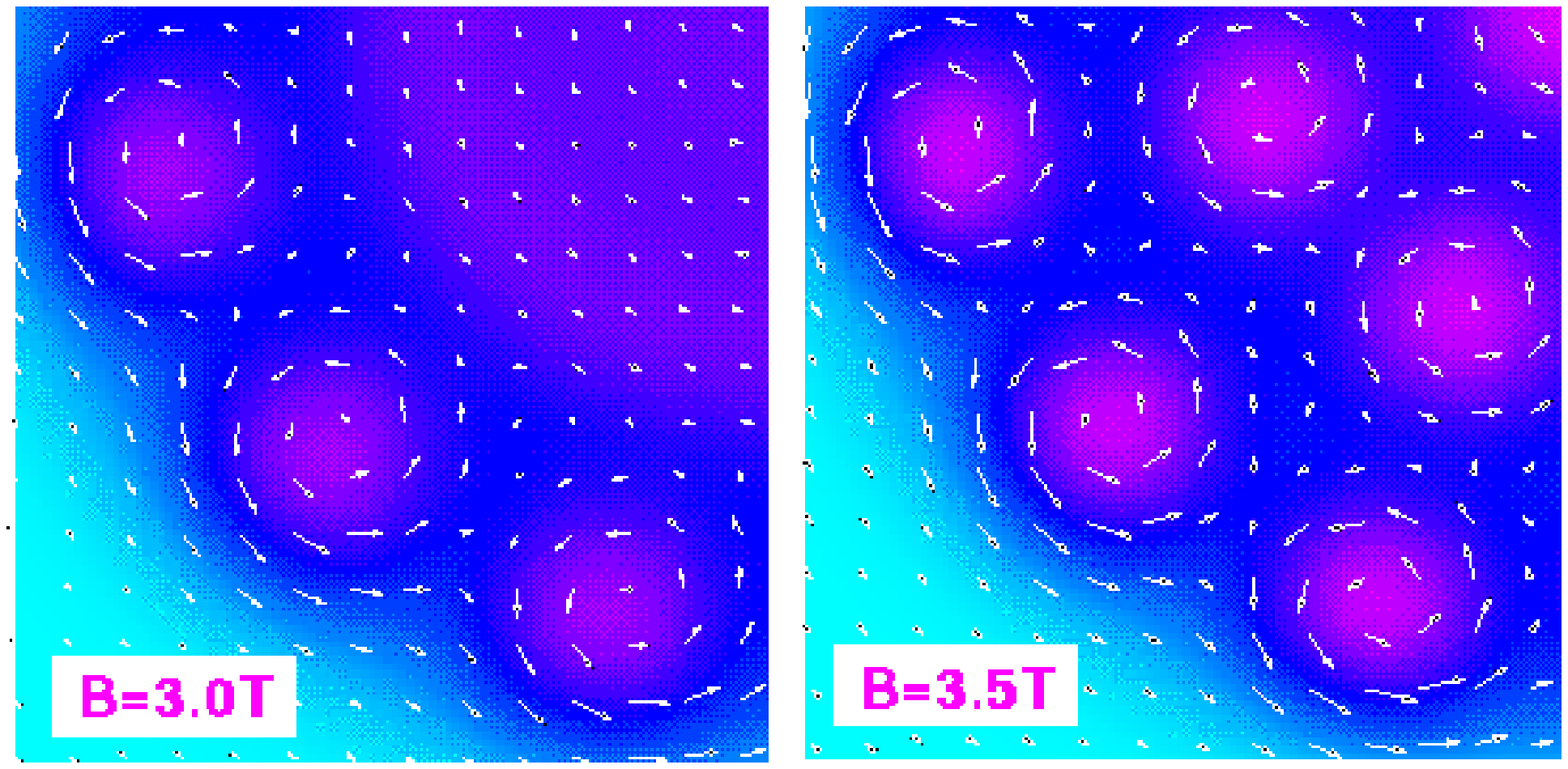}{0.5}{14}{\normalsize 
Real particle current ${\bf j}({\bf r})$ 
for $N=20$ at $B=3.0$T ({\it left}) and $B=3.5$T ({\it right}),
shown in the lower left quadrant of the charge density.
The maximum currents are about 0.028(nm ps)$^{-2}$ (at 3.0T) and 
0.035(nm ps)$^{-2}$ (at 3.5T). The shaded background indicates
the charge density distribution: areas colored in magenta correspond to 
the regions with largest charge density
}{fig2}
For still higher fields, rotational symmetry is also broken 
for the inner parts of the droplet (see lowest panel in Fig.~\ref{fig1}). 
The reconstructed density now forms a {\it sequence of rings},
each consisting of well-separated maxima in the electron density.
Correspondingly, inside the droplet the real
current ${\bf j}(x,y)$ forms vortices (see right panel of Fig.~\ref{fig2})
centered at the separate density maxima along the rings. 
The localization of all electrons is associated with opening a large 
gap at the Fermi level. This had not been the case if only electrons 
along the edge were localized, as the inner part of the droplet 
then still was in the MDD phase.
In the high-field limit, the  ``bumpy'' electron density 
is consistent with the numerical results of HF calculations by  
M\"uller and Koonin~\cite{koonin}. Performing exact calculations for
$N\le 6$, Maksym~\cite{maksym}
found localized states using a rotating frame.
\PostScript{3.2}{0}{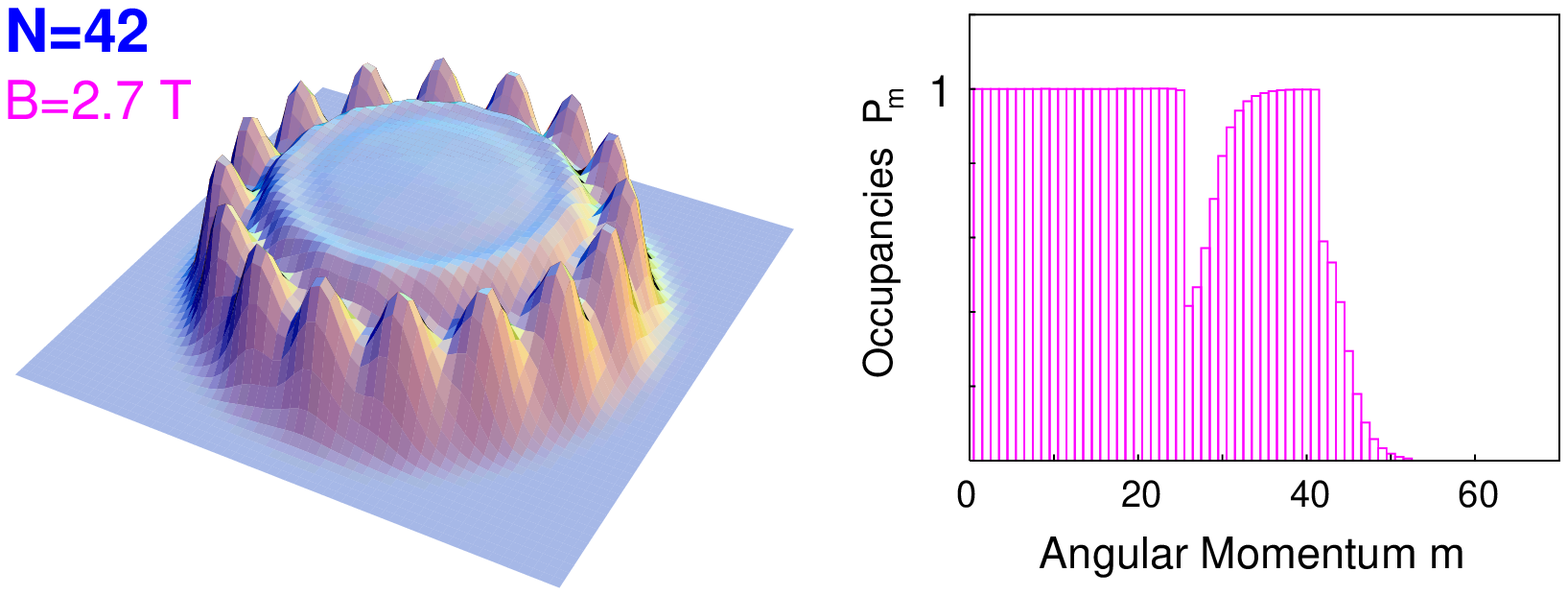}{0.5}{14}{
As Fig.~\ref{fig1}, but for $N=42$ electrons and at a magnetic field $B=2.7$T
}{fig3}
The broken symmetry of the Chamon-Wen edge is not limited to small dots.
Figure~\ref{fig3} shows the density and angular momentum distribution
of a dot with $N=42$ electrons. 
Again the edge consists of nearly localized electrons.
The inner MDD at filling factor one is very stable, with integer 
occupancy of angular momentum.
By analyzing the Kohn-Sham single-particle orbitals we noticed that they 
fall into two discrete subsets, one forming the MDD and the other 
forming the broken-symmetry Chamon-Wen edge. This opens up a possibility 
for collective excitations localized at the edge.
\PostScript{6}{0}{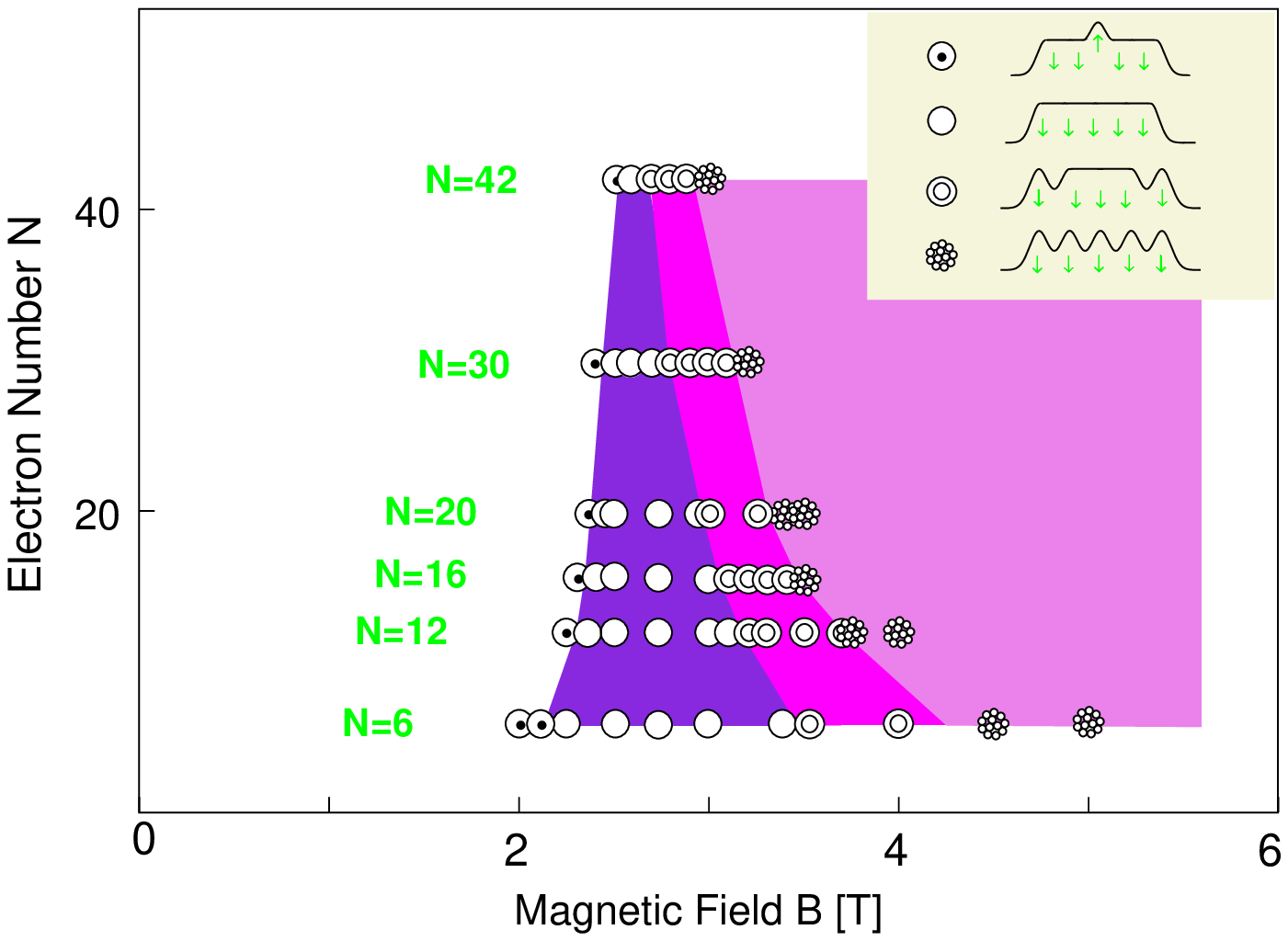}{0.5}{14}{\normalsize 
Different phases for a quantum dot with $N=6, 12, 16, 20, 30$ 
and $N=42$ electrons at a density roughly corresponding to $r_s=2 a_B^*$.
The colored areas indicate regions of magnetic fields where 
different phases (polarization, MDD, reconstruction and beyond) occur:
The purple region indicates the MDD phase, while 
the magenta region corresponds to the Chamon-Wen edge regime.
In the pink region localization extends over the whole droplet 
}{fig4}
We finally 
study the formation of the MDD and its reconstruction  
systematically as a function of the electron number.
The resulting phase diagram is shown in Fig.~\ref{fig4}.
We find three phase boundaries: the polarization transition with the 
subsequent formation of the MDD, the broken-symmetry Chamon-Wen edge phase
and finally the localization phase~\cite{koonin} where the 
MDD completely disappears.
The range of magnetic fields where the 
MDD is a stable ground-state becomes much more narrow with increasing 
$N$, as it has been noted in Refs.~\cite{mcdonald,cw,ferconi}. 
The phase boundaries are in good agreement with recent experiments 
by Oosterkamp~{\it et al.}~\cite{oosterkamp}. (The fact that we find the 
transitions between different phases 
at smaller fields is due to a different electron 
density~\cite{foot2}). 
The experimental results also show systematic changes in the chemical potential
after the first appearance of the Chamon-Wen edge. 
This might be related to the formation of rings and the stepwise 
disappearance of the MDD, which has a similar dependence of 
the number of electrons and the magnetic field.

We have exclusively investigated
the reconstruction in the regime where the droplet was completely polarized
and refer to~\cite{karlhede,franco} for
a discussion about the existence of spin textures along the edges. 
Performing  fully unrestricted Hartree-Fock calculations, 
Karlhede and Lejnell~\cite{karlhede} as well as Franco and Brey~\cite{franco}
found charge density wave-like modulations along the reconstructed edge 
of a Hall bar. These results are similar to our finding of the broken-symmetry 
edge in {\it finite} droplets.

In conclusion, we found that the maximum density droplet 
reconstructs into {\it Chamon-Wen edge states with broken rotational symmetry}
in the internal coordinates.
For larger fields reconstruction continues by sequential formation 
of rings up to an overall bumpy ground state density. 
The transitions with increasing $B$ define different phases 
of the electron liquid inside the droplet. We showed a phase diagram
which displayed 
a systematic separation between different phases of reconstruction 
both as a function of dot size and magnetic field.
The trends for the shape of the phase boundaries between 
polarization transition, MDD, broken symmetry edge reconstruction and finally 
localized states~\cite{koonin} are rather clear and
agree~\cite{foot2} with the phase diagram recently measured by 
Oosterkamp~{\it et al.}~\cite{oosterkamp}.

\bigskip

We acknowledge discussions with D. G. Austing and S. Viefers. 
This work was supported by the Academy of Finland, the Studienstiftung 
des deutschen Volkes and the TMR programme of the European 
Community under contract ERBFMBICT972405.

\end{multicols}

\begin{thebibliography}{99}
\bibitem{foot4} Note that we omit the negative sign of angular momentum 
eigenvalues here and in the following.
\end{thebibliography}
\end{document}